\documentclass[onecolumn]{revtex4}
\usepackage{amsfonts}
\usepackage{amsmath}

\begin{document}

\title{New approach for deriving operator identities by alternately using normally,
antinormally, and Weyl ordered integration technique\thanks{Supported by the
National Natural Science Foundation of China under grant Nos.10775097 and
10874174.}}
\author{Hong-Yi Fan and Hong-Chun Yuan\thanks{Corresponding author, E-mail:
yuanhch@sjtu.edu.cn.}\\{\small Department of Physics, Shanghai Jiao Tong University, Shanghai,
200030, P. R. China}}

\begin{abstract}
Dirac's ket-bra formalism is the "language" of quantum mechanics and quantum
field theory. In Ref.\cite{r1,r2} (Fan et al, Ann. Phys. 321 (2006) 480; 323
(2008) 500) we have reviewed how to apply Newton-Leibniz integration rules to
Dirac's ket-bra projectors. In this work by alternately using the technique of
integration within normal, antinormal, and Weyl ordering of operators we not
only derive some new operator ordering identities, but also deduce some useful
integration formulas regarding to Laguerre and Hermite polynomials. This opens
a new route of deriving mathematical integration formulas by virtue of the
quantum mechanical operator ordering technique.


\end{abstract}
\maketitle

\section{Introduction}

In the foreword of the book $\ll$The Principles of Quantum Mechanics$\gg$
Dirac wrote: "\textit{The symbolic method, }$\cdots$\textit{enables one to
express the physical law in a neat and concise way, and will probably be
increasingly used in the future as it becomes better understood and its own
special mathematics gets developed.}"\cite{r3} Following his expectation, the
technique of integration within an ordered product (IWOP) of operators was
invented which can directly apply Newton-Leibniz integration rule to ket-bra
projective operators. \cite{r1,r2} The essence of IWOP technique is to convert
these ket-bra operators into certain ordered product (normal ordering,
antinormal ordering, and Weyl ordering) of bosonic creation and annihilation
operators such that they can be treated as ordinary parameters while
performing the integrations, but operators' essence will not lose in this
approach. For example, using the IWOP technique\cite{r4,r5} we have converted
the completeness relation of the coordinate eigenvector $\left \vert
x\right \rangle $ as a pure Gaussian integration within the normal ordering
$\colon \colon$,
\begin{equation}
\int_{-\infty}^{\infty}dx\left \vert x\right \rangle \left \langle x\right \vert
=\int_{-\infty}^{\infty}\frac{dx}{\sqrt{\pi}}\colon e^{-\left(  x-X\right)
^{2}}\colon=1,\text{ }\label{1}%
\end{equation}
where $X=(a+a^{\dagger})/\sqrt{2}$ is the coordinate operator with $\left[
a,a^{\dagger}\right]  =1$. It turns out that performing the ket-bra
integration $\int_{-\infty}^{\infty}\frac{dx}{\mu}\left \vert \frac{x}{\mu
}\right \rangle \left \langle x\right \vert $ leads to the single-mode squeezing
operator. Recently, we directly and concisely obtain the connection between
Wick ordered polynomial and Hermite polynomials\cite{r5-0} by virtue of the
the IWOP technique.\cite{r5-1}

In this work we shall demonstrate how to derive some new operator ordering
identities and new integration formulas regarding to Laguerre polynomials
$L_{n}^{\left(  \alpha \right)  }\left(  x\right)  $ and Hermite polynomials
$H_{n}\left(  x\right)  $ by alternately using the technique of integration
within normal, antinormal, and Weyl ordering of operators. This opens a new
route of deriving mathematical integration formulas by virtue of the quantum
mechanical operator ordering technique.

\section{Antinormally ordered expansion of the operator $X^{n}$ and
$H_{n}\left(  X\right)  $}

To begin with, using the Baker-Hausdorff formula
\begin{equation}
e^{A+B}=e^{A}e^{B}e^{-\frac{1}{2}\left[  A,B\right]  }=e^{B}e^{A}e^{-\frac
{1}{2}\left[  B,A\right]  } \label{a1}%
\end{equation}
and the generating function of $H_{n}\left(  x\right)  $%
\begin{equation}
\sum_{n=0}^{\infty}\frac{t^{n}}{n!}H_{n}\left(  x\right)  =e^{2tx-t^{2}}
\label{a2-0}%
\end{equation}
where%
\begin{equation}
H_{n}\left(  x\right)  =\sum_{l=0}^{\left[  n/2\right]  }\frac{\left(
-1\right)  ^{l}n!}{l!(n-2l)!}(2x)^{n-2l},
\end{equation}
we easily obtain%

\begin{equation}
e^{\lambda X}=e^{\frac{\lambda}{\sqrt{2}}\left(  a+a^{\dagger}\right)
}=\vdots e^{\lambda X-\frac{1}{4}\lambda^{2}}\vdots=\sum_{n=0}^{\infty}%
\frac{\left(  \frac{\lambda}{2}\right)  ^{n}}{n!}\vdots H_{n}\left(  X\right)
\vdots, \label{a3}%
\end{equation}
where $\vdots \vdots$ stands for antinormal ordering. Comparing Eq.(\ref{a3})
with the same power of $\lambda$ in the expansion of $e^{\lambda X}%
=\sum \limits_{n=0}^{\infty}\frac{\lambda^{n}}{n!}X^{n}$, we have the neat
formula
\begin{equation}
X^{n}=2^{-n}\vdots H_{n}\left(  X\right)  \vdots, \label{a4}%
\end{equation}
which is just the antinormal ordering expansion of $X^{n}$. Taking $n=2$ for
example,
\begin{align}
\frac{1}{4}\vdots H_{2}\left(  X\right)  \vdots &  =\frac{1}{4}\vdots \left(
4X^{2}-2\right)  \vdots \nonumber \\
&  =\frac{1}{2}\left(  a^{2}+2aa^{\dagger}+a^{\dagger2}-1\right)  =X^{2},
\end{align}
as expected. It then follows from Eq.(\ref{a4}) that%

\begin{equation}
X^{n+m}=2^{-n-m}\vdots H_{n+m}\left(  X\right)  \vdots=X^{n}X^{m}, \label{a5}%
\end{equation}
which implies the following new relation%

\begin{equation}
\vdots H_{n+m}\left(  X\right)  \vdots=\vdots H_{n}\left(  X\right)
\vdots \vdots H_{m}\left(  X\right)  \vdots. \label{a6}%
\end{equation}
From the relation
\begin{equation}
\frac{d}{dx}H_{n}\left(  x\right)  =2nH_{n-1}\left(  x\right)  , \label{a6-1}%
\end{equation}
and Eq.(\ref{a4}) we have%
\begin{equation}
\frac{d}{dX}\vdots H_{n}\left(  X\right)  \vdots=2^{n}nX^{n-1}=2n\vdots
H_{n-1}\left(  X\right)  \vdots=\vdots \frac{d}{dX}H_{n}\left(  X\right)
\vdots, \label{a6-2}%
\end{equation}
namely,
\begin{equation}
\frac{d}{dX}\vdots H_{n}\left(  X\right)  \vdots=\vdots \frac{d}{dX}%
H_{n}\left(  X\right)  \vdots. \label{a6-3}%
\end{equation}
This is another property of $\vdots H_{n}\left(  X\right)  \vdots$.

Moreover, from Eqs.(\ref{a2-0}) and (\ref{a1}) we see%
\begin{align}
\sum_{n=0}^{\infty}\frac{t^{n}}{n!}H_{n}\left(  \lambda X\right)   &
=e^{2t\lambda X-t^{2}}=\vdots e^{2t\lambda X-\left(  \lambda^{2}+1\right)
t^{2}}\vdots \nonumber \\
&  =\sum_{n=0}^{\infty}\frac{\left(  \sqrt{\left(  \lambda^{2}+1\right)
}t\right)  ^{n}}{n!}\vdots H_{n}\left(  \frac{\lambda X}{\sqrt{\left(
\lambda^{2}+1\right)  }}\right)  \vdots \label{a7}%
\end{align}
Comparing the same power of $t$ on the two sides we obtain a new identity%
\begin{equation}
H_{n}\left(  \lambda X\right)  =\left(  \sqrt{\left(  \lambda^{2}+1\right)
}\right)  ^{n}\vdots H_{n}\left(  \frac{\lambda X}{\sqrt{\left(  \lambda
^{2}+1\right)  }}\right)  \vdots \label{a7-1}%
\end{equation}
Especially, when $\lambda=1$, Eq.(\ref{a7-1}) reduces to
\begin{equation}
H_{n}\left(  X\right)  =\left(  \sqrt{2}\right)  ^{n}\vdots H_{n}\left(
\frac{X}{\sqrt{2}}\right)  \vdots. \label{a8}%
\end{equation}
which is different from Eq.(\ref{a4}). For $n=2$ case,%
\begin{align}
2\vdots H_{2}\left(  \frac{X}{\sqrt{2}}\right)  \vdots &  =\vdots2\left(
a+a^{\dagger}\right)  ^{2}-4\vdots \nonumber \\
&  =2\left(  a^{2}+2aa^{\dagger}+a^{\dagger2}\right)  -4\nonumber \\
&  =4\left(  \frac{a+a^{\dagger}}{\sqrt{2}}\right)  ^{2}-2\nonumber \\
&  =H_{2}\left(  X\right)
\end{align}

By analogy to the above derivation we can deduce the normal ordering expansion
of $X^{n}$
\begin{equation}
X^{n}=\left(  2i\right)  ^{-n}\colon H_{n}\left(  iX\right)  \colon,
\label{001}%
\end{equation}
and the normally ordered form of $H_{n}\left(  X\right)  $%
\begin{equation}
H_{n}\left(  X\right)  =2^{n}\colon X^{n}\colon. \label{002}%
\end{equation}

\section{New integration formulas about Hermite polynomials}

Using the above results we can derive new integration formulas regarding to
Hermite polynomials (HP) $H_{n}\left(  x\right)  $. Recalling the
$P$-representation in the coherent state $\left \vert \beta \right \rangle $
basis\cite{r6}
\begin{equation}
\rho=\int \frac{d^{2}\beta}{\pi}P\left(  \beta \right)  \left \vert
\beta \right \rangle \left \langle \beta \right \vert ,\beta \equiv \beta_{1}%
+i\beta_{2} \label{03}%
\end{equation}
where
\begin{equation}
\left \vert \beta \right \rangle =\exp \left[  -\frac{\left \vert \beta \right \vert
^{2}}{2}+\beta a^{\dagger}\right]  \left \vert 0\right \rangle , \label{04}%
\end{equation}
is the coherent state satisfying with
\begin{equation}
a\left \vert \beta \right \rangle =\beta \left \vert \beta \right \rangle .
\label{04-1}%
\end{equation}
Utilizing Eqs.(\ref{a4}) and (\ref{03}) as well as the vacuum projector
$\left \vert 0\right \rangle \left \langle 0\right \vert =\colon \exp \left(
-a^{\dagger}a\right)  \colon$and considering (\ref{001}), we have
\begin{align}
X^{n}  &  =2^{-n}\int \frac{d^{2}\beta}{\pi}H_{n}\left(  \frac{\beta
+\beta^{\ast}}{\sqrt{2}}\right)  \left \vert \beta \right \rangle \left \langle
\beta \right \vert \nonumber \\
&  =2^{-n}\int \frac{d^{2}\beta}{\pi}H_{n}\left(  \sqrt{2}\beta_{1}\right)
\colon \exp \left(  -\left \vert \beta \right \vert ^{2}+\beta a^{\dagger}%
+\beta^{\ast}a-a^{\dagger}a\right)  \colon \nonumber \\
&  =\left(  2i\right)  ^{-n}\colon H_{n}\left(  iX\right)  \colon. \label{05}%
\end{align}
Since within the normal ordering symbol $a$ and $a^{\dagger}$ are commute,
they can be treated as $c$-numbers. So we can set $a\rightarrow x$ and
$a^{\dagger}\rightarrow y$, in this way we obtain%
\begin{equation}
\int \frac{d^{2}\beta}{\pi}H_{n}\left(  \sqrt{2}\beta_{1}\right)  \exp \left(
-\left \vert \beta \right \vert ^{2}+y\beta+x\beta^{\ast}\right)  =i^{-n}%
H_{n}\left(  i\frac{x+y}{\sqrt{2}}\right)  e^{xy}, \label{07}%
\end{equation}
which is a new integration formula. Here we derive it without really
performing the integration in Eq.(\ref{07}). This is an obvious advantage of
the IWOP technique.

Next, the antinormally ordered expansion of an arbitrary bosonic operator in
the coherent state $\left \vert \beta \right \rangle $ basis is\cite{r7}
\begin{equation}
\rho=\int \frac{d^{2}\beta}{\pi}\vdots \left \langle -\beta \right \vert
\rho \left \vert \beta \right \rangle \exp[\left \vert \beta \right \vert ^{2}%
+\beta^{\ast}a-\beta a^{\dagger}+a^{\dagger}a]\vdots \label{003}%
\end{equation}
provided that the integral is convergent, which shows that when $\rho$ is in
normal ordering, its coherent state matrix element $\left \langle
-\beta \right \vert \rho \left \vert \beta \right \rangle $ can be immediately
obtained. So we can get $\rho$'s antinormally ordered expansion by just
performing the integral of Eq.(\ref{003}). According to Eqs.(\ref{001}) and
(\ref{003}) and considering Eq.(\ref{a4}) again, we obtain the following
equation%
\begin{align}
X^{n}  &  =\left(  2i\right)  ^{-n}\int \frac{d^{2}\beta}{\pi}\vdots
\left \langle -\beta \right \vert \colon H_{n}\left(  iX\right)  \colon \left \vert
\beta \right \rangle \exp[\left \vert \beta \right \vert ^{2}+\beta^{\ast}a-\beta
a^{\dagger}+a^{\dagger}a]\vdots \nonumber \\
&  =\left(  2i\right)  ^{-n}\int \frac{d^{2}\beta}{\pi}\vdots H_{n}\left(
-\sqrt{2}\beta_{2}\right)  \exp[-\left \vert \beta \right \vert ^{2}+\beta^{\ast
}a-\beta a^{\dagger}+a^{\dagger}a]\vdots \nonumber \\
&  =2^{-n}\vdots H_{n}\left(  X\right)  \vdots. \label{004}%
\end{align}
Thus, one immediately gets another new integration formula%
\begin{equation}
\int \frac{d^{2}\beta}{\pi}H_{n}\left(  -\sqrt{2}\beta_{2}\right)
\exp[-\left \vert \beta \right \vert ^{2}+x\beta^{\ast}-\beta y]=i^{n}%
H_{n}\left(  \frac{x+y}{\sqrt{2}}\right)  e^{-xy}, \label{006}%
\end{equation}
which is different from Eq.(\ref{07}).

On the other hand, in Refs.\cite{r8,r8-1,r8-2}, by considering $P$%
-representation of an operator $\rho$ and using the Weyl ordering of the
coherent state projector%
\begin{equation}
\left \vert \beta \right \rangle \left \langle \beta \right \vert =2%
\begin{array}
[c]{c}%
\colon \\
\colon
\end{array}
\exp \left[  -2\left(  \beta^{\ast}-a^{\dagger}\right)  \left(  \beta-a\right)
\right]
\begin{array}
[c]{c}%
\colon \\
\colon
\end{array}
, \label{007}%
\end{equation}
we have derived the useful formula which can convert $\rho$ into its Weyl
ordered form%

\begin{equation}
\rho=2\int \frac{d^{2}\beta}{\pi}%
\begin{array}
[c]{c}%
\colon \\
\colon
\end{array}
\left \langle -\beta \right \vert \rho \left \vert \beta \right \rangle \exp \left[
2\left(  \beta^{\ast}a-a^{\dagger}\beta+a^{\dagger}a\right)  \right]
\begin{array}
[c]{c}%
\colon \\
\colon
\end{array}
, \label{008}%
\end{equation}
where $%
\begin{array}
[c]{c}%
\colon \colon \\
\colon \colon
\end{array}
$ stands for Weyl ordering, $a$ and $a^{\dagger}$ are commute within $%
\begin{array}
[c]{c}%
\colon \colon \\
\colon \colon
\end{array}
$. By using Eqs.(\ref{002}) and (\ref{008}), we obtain%

\begin{align}
H_{n}\left(  X\right)   &  =2^{n+1}\int \frac{d^{2}\beta}{\pi}%
\begin{array}
[c]{c}%
\colon \\
\colon
\end{array}
\left \langle -\beta \right \vert \colon X^{n}\colon \left \vert \beta \right \rangle
\exp \left[  2\left(  \beta^{\ast}a-a^{\dagger}\beta+a^{\dagger}a\right)
\right]
\begin{array}
[c]{c}%
\colon \\
\colon
\end{array}
\nonumber \\
&  =i^{n}2^{\frac{3n+2}{2}}\int \frac{d^{2}\beta}{\pi}%
\begin{array}
[c]{c}%
\colon \\
\colon
\end{array}
\beta_{2}^{n}\exp \left[  2\left(  -|\beta|^{2}+\beta^{\ast}a-a^{\dagger}%
\beta+a^{\dagger}a\right)  \right]
\begin{array}
[c]{c}%
\colon \\
\colon
\end{array}
. \label{009}%
\end{align}
Due to
\begin{equation}
H_{n}\left(  X\right)  =%
\begin{array}
[c]{c}%
\colon \\
\colon
\end{array}
H_{n}\left(  X\right)
\begin{array}
[c]{c}%
\colon \\
\colon
\end{array}
,
\end{equation}
we have a new integration formula%
\begin{equation}
i^{n}2^{\frac{3n+2}{2}}\int \frac{d^{2}\beta}{\pi}\beta_{2}^{n}\exp \left[
2\left(  -|\beta|^{2}+x\beta^{\ast}-y\beta \right)  \right]  =H_{n}\left(
\frac{x+y}{\sqrt{2}}\right)  e^{-2xy}, \label{010}%
\end{equation}
without really performing this integration.

Finally, in the similar way for the momentum $P=\frac{1}{i\sqrt{2}%
}(a+a^{\dagger})$, we easily obtain some integration formula as follows
\begin{equation}
\int \frac{d^{2}\beta}{\pi}H_{n}\left(  \sqrt{2}\beta_{2}\right)  \exp \left(
-\left \vert \beta \right \vert ^{2}+y\beta+x\beta^{\ast}\right)  =i^{-n}%
H_{n}\left(  \frac{x-y}{\sqrt{2}}\right)  e^{xy},
\end{equation}%
\begin{equation}
\int \frac{d^{2}\beta}{\pi}H_{n}\left(  \sqrt{2}\beta_{1}\right)  \exp \left(
-\left \vert \beta \right \vert ^{2}+x\beta^{\ast}-y\beta \right)  =i^{n}%
H_{n}\left(  \frac{x-y}{i\sqrt{2}}\right)  e^{-xy},
\end{equation}
and%
\begin{equation}
\left(  -i\right)  ^{n}2^{\frac{3n+2}{2}}\int \frac{d^{2}\beta}{\pi}\beta
_{1}^{n}\exp \left[  2\left(  -|\beta|^{2}+\beta^{\ast}a-a^{\dagger}%
\beta \right)  \right]  =H_{n}\left(  \frac{x-y}{i\sqrt{2}}\right)  e^{-2xy}.
\end{equation}

\section{Operator identities about\ Laguerre polynomials}

Using Eq.(\ref{1}) and the IWOP technique, we have the normally ordered
expansion formula%

\begin{align}
e^{\lambda X^{2}}  &  =\int_{-\infty}^{\infty}dxe^{\lambda x^{2}}\left \vert
x\right \rangle \left \langle x\right \vert =\colon \int_{-\infty}^{\infty}%
\frac{dx}{\sqrt{\pi}}e^{\lambda x^{2}}e^{-\left(  x-X\right)  ^{2}}%
\colon \nonumber \\
&  =\left(  1-\lambda \right)  ^{-1/2}\colon \exp[\frac{-\lambda X^{2}}%
{\lambda-1}]\colon. \label{d1}%
\end{align}
Comparing Eq.(\ref{d1}) with the generating function of the Laguerre
polynomials%
\begin{equation}
\left(  1-t\right)  ^{-\alpha-1}\exp[\frac{xt}{t-1}]=\sum_{n=0}^{\infty}%
L_{n}^{\alpha}\left(  x\right)  t^{n}, \label{d2}%
\end{equation}
we see when $\alpha=-1/2,$%
\begin{equation}
\colon \sum_{n=0}^{\infty}L_{n}^{-1/2}\left(  -X^{2}\right)  \lambda^{n}%
\colon=e^{\lambda X^{2}}, \label{d3}%
\end{equation}
which is a new operator identity. Then comparing Eq.(\ref{d3}) with
$e^{\lambda X^{2}}=\sum_{n=0}^{\infty}\frac{\lambda^{n}}{n!}X^{2n}$, we see%
\begin{equation}
X^{2n}=n!\colon L_{n}^{-1/2}\left(  -X^{2}\right)  \colon. \label{d4}%
\end{equation}
Due to $X^{2n}=\left(  2i\right)  ^{-2n}\colon H_{2n}\left(  iX\right)
\colon$in Eq.(\ref{001}), we obtain the relation connecting Hermite polynomial
and Laguerre polynomial
\begin{equation}
H_{2n}\left(  iX\right)  =\left(  -1\right)  ^{n}2^{2n}n!L_{n}^{-1/2}\left(
-X^{2}\right)  , \label{d5}%
\end{equation}
this relation still holds when $iX\rightarrow x$, i.e.,%
\begin{equation}
H_{2n}\left(  x\right)  =\left(  -1\right)  ^{n}2^{2n}n!L_{n}^{-1/2}\left(
x^{2}\right)  , \label{d6}%
\end{equation}
which coincides with Ref.\cite{r9}. Since $H_{2n}\left(  X\right)
=2^{2n}\colon X^{2n:}\colon$in Eq.(\ref{002}), we also have%
\begin{equation}
\colon X^{2n}\colon=\left(  -1\right)  ^{n}n!L_{n}^{-1/2}\left(  X^{2}\right)
. \label{d7}%
\end{equation}

Considering Eqs.(\ref{003}) and (\ref{d4}), we derive%
\begin{align}
X^{2n}  &  =n!\int \frac{d^{2}\beta}{\pi}\vdots \left \langle -\beta \right \vert
\colon L_{n}^{-1/2}\left(  -X^{2}\right)  \colon \left \vert \beta \right \rangle
\exp[\left \vert \beta \right \vert ^{2}+\beta^{\ast}a-\beta a^{\dagger
}+a^{\dagger}a]\vdots \nonumber \\
&  =n!\int \frac{d^{2}\beta}{\pi}\vdots L_{n}^{-1/2}\left(  2\beta_{2}%
^{2}\right)  \exp[-\left \vert \beta \right \vert ^{2}+\beta^{\ast}a-\beta
a^{\dagger}+a^{\dagger}a]\vdots. \label{d8}%
\end{align}
It then follows from Eq.(\ref{a4}) that%
\begin{equation}
n!\int \frac{d^{2}\beta}{\pi}L_{n}^{-1/2}\left(  2\beta_{2}^{2}\right)
\vdots \exp[-\left \vert \beta \right \vert ^{2}+\beta^{\ast}a-\beta a^{\dagger
}+a^{\dagger}a]\vdots=2^{-2n}\vdots H_{2n}\left(  X\right)  \vdots, \label{d9}%
\end{equation}
which is a new identity. Further, when $a\rightarrow x$ and $a^{\dagger
}\rightarrow y$, we deduce a new integration formula as follows%
\begin{equation}
n!\int \frac{d^{2}\beta}{\pi}L_{n}^{-1/2}\left(  2\beta_{2}^{2}\right)
\exp[-\left \vert \beta \right \vert ^{2}+x\beta^{\ast}-y\beta]=2^{-2n}%
H_{2n}\left(  \frac{x+y}{\sqrt{2}}\right)  e^{-xy}. \label{d10}%
\end{equation}

In the similar manner, by considering Eqs.(\ref{008}) and (\ref{d7}), we can
obtain%
\begin{align}
\left(  -1\right)  ^{n}n!L_{n}^{-1/2}\left(  X^{2}\right)   &  =2\int
\frac{d^{2}\beta}{\pi}%
\begin{array}
[c]{c}%
\colon \\
\colon
\end{array}
\left \langle -\beta \right \vert \colon X^{2n}\colon \left \vert \beta
\right \rangle \exp \left[  2\left(  \beta^{\ast}a-a^{\dagger}\beta+a^{\dagger
}a\right)  \right]
\begin{array}
[c]{c}%
\colon \\
\colon
\end{array}
\nonumber \\
&  =\left(  -1\right)  ^{n}2^{n+1}\int \frac{d^{2}\beta}{\pi}%
\begin{array}
[c]{c}%
\colon \\
\colon
\end{array}
\beta_{2}^{2n}\exp \left[  2\left(  -\left \vert \beta \right \vert ^{2}%
+\beta^{\ast}a-a^{\dagger}\beta+a^{\dagger}a\right)  \right]
\begin{array}
[c]{c}%
\colon \\
\colon
\end{array}
.
\end{align}
From the above equation, due to $L_{n}^{-1/2}\left(  X^{2}\right)  =%
\begin{array}
[c]{c}%
\colon \\
\colon
\end{array}
L_{n}^{-1/2}\left(  X^{2}\right)
\begin{array}
[c]{c}%
\colon \\
\colon
\end{array}
$, we see that
\begin{equation}
2^{n+1}\int \frac{d^{2}\beta}{\pi}\beta_{2}^{2n}%
\begin{array}
[c]{c}%
\colon \\
\colon
\end{array}
\exp \left[  2\left(  -\left \vert \beta \right \vert ^{2}+\beta^{\ast
}a-a^{\dagger}\beta+a^{\dagger}a\right)  \right]
\begin{array}
[c]{c}%
\colon \\
\colon
\end{array}
=n!%
\begin{array}
[c]{c}%
\colon \\
\colon
\end{array}
L_{n}^{-1/2}\left(  X^{2}\right)
\begin{array}
[c]{c}%
\colon \\
\colon
\end{array}
.
\end{equation}
It then follows the new integration formula
\begin{equation}
2^{n+1}\int \frac{d^{2}\beta}{\pi}\beta_{2}^{2n}\exp \left[  2\left(
-\left \vert \beta \right \vert ^{2}+x\beta^{\ast}-y\beta \right)  \right]
=n!L_{n}^{-1/2}\left[  \frac{1}{2}\left(  x+y\right)  ^{2}\right]  e^{-2xy}.
\end{equation}

In summary, we have introduced an effective approach for deriving operator
identities and new integration formulas by alternately using normally,
antinormally, and Weyl ordered integration technique. This opens a new route
of deriving mathematical integration formulas by virtue of the quantum
mechanical operator ordering technique.

\end{document}